\documentclass[12pt]{article}
\usepackage{amsmath,amssymb,amsfonts}
\usepackage[dvips]{graphicx}
\usepackage{epsfig}
\usepackage{color}

\newcommand{\be}{\begin{equation}}
\newcommand{\ee}{\end{equation}}
\newcommand{\ba}{\begin{eqnarray}}
\newcommand{\ea}{\end{eqnarray}}
\newcommand{\bea}{\begin{array}}
\newcommand{\eea}{\end{array}}

\newcommand{\CL}{{\cal L}}

\newcommand{\Tr}{{\rm Tr}}
\newcommand{\nn}{{\nonumber}}

\makeatletter \@addtoreset{equation}{section} \makeatother

\begin{document}

\begin{titlepage}
\vfill
\begin{flushright}
{\tt\normalsize KIAS-P09001}\\
\end{flushright}
\vfill
\begin{center}
{\large\bf    Janus and Multifaced Supersymmetric  Theories II}

\vfill
Chanju Kim$^{1,3}$, Eunkyung Koh$^2$, and Ki-Myeong Lee$^3$,

\vskip 2mm
$^1${\it Dept. of Physics, Ewha Womans University, Seoul 120-750, Korea} \\
\vskip1mm
$^2${\it Dept. of Physics, Seoul National University,
Seoul 151-747, Korea}
\\
\vskip 1mm
$^3${\it   Korea Institute for Advanced Study,
Seoul 130-722, Korea}
\vfill
\end{center}

\begin{abstract}
\noindent  We explore the physics of supersymmetric Janus gauge theories
in four dimensions with spatial dependent  coupling constants  $e^2$ and $\theta$. 
For the 8 supersymmetric case, we 
study the vacuum and Bogomol'nyi-Prasad-Sommerfield spectrum, and the physics of a sharp interface where
the couple constants jump. We also find less supersymmetric cases either due to
additional expressions in the Lagrangian or to the fact that coupling constants depend
on additional spatial coordinates.

\end{abstract}

\vfill
\end{titlepage}

\section{Introduction and Conclusion}

There has been some interest in the Janus-type field theory
where the coupling constants are dependent on space~\cite{D'Hoker:2006uv}. What is exciting about
this subject is that some of supersymmetry of the original theory can be
preserved even when the coupling constants are spatially  dependent
once some corrections are made on the Lagrangian and the supersymmetric transformation.
Such field theories appear naturally in  the context of  AdS-geometry~\cite{Bak:2003jk}.
Recently, a Janus-type field theory has been discovered in the intersecting D3-$(p,q)$ 5 brane~\cite{Gaiotto:2008sd,Gaiotto:2008ak}. The field theory on D3 branes can have a spatially dependent  complex
coupling    $\tau=\theta/2\pi + 4\pi i/e^2$, preserving half of original 16 supersymmetries.

In this work we explore further these theories and generalize them, extending our previous work~\cite{Kim:2008dj}.
  We first study the vacuum and Bomomol'yni-Prasad-Sommerfield (BPS)
configurations. We also study  the wave and  dyonic physics near  a sharp interface, 
which acts like an axionic domain wall. We also find additional supersymmetry
 breaking Janus theories.

 The original Janus solution in
Ref.~\cite{Bak:2003jk} is a 1-parameter family of dilatonic
deformations of $AdS_5$ space without supersymmetry.    The
Janus solution is made of  two Minkowski spaces joined along an
interface so that the dilaton field interpolates two asymptotic
values at two spaces. The CFT dual field theory is suggested to be
the deformation of the  Yang-Mills theory where  the coupling
constant changes from one region to another
region~\cite{Bak:2003jk}.

Further works revealed that one can have  supersymmetric Janus
geometries with the various supersymmetries and internal
symmetries\cite{Clark:2004sb,Clark:2005te,D'Hoker:2006uu,Gomis:2006cu,D'Hoker:2007xy}. Starting from the 16
supersymmetric Yang-Mills theory, the various deformations of  
2, 4, 8  supersymmetries  have  been found\cite{D'Hoker:2006uv}.
 Especially, the 16 supersymmetric Janus geometries
have been found~\cite{Gomis:2006cu,D'Hoker:2007xy}, where both dilaton and axion fields vary along a spatial direction. Also, other aspects of the
Janus solutions have been discussed in 
Refs.~\cite{Bak:2004yf,Sonner:2005sj,Bak:2006nh,Hirano:2006as,Bak:2007jm}.

In our previous work~\cite{Kim:2008dj}, we investigated in detail  the vacuum and BPS structure of
the supersymmetric Janus theory for the case where only $e^2$ depends on the spatial coordinates and found that there can be a new type of classical vacua, which are
characterized by the Nahm equation when there are planes where the coupling constant $e^2$ vanishes. 
In a later work~\cite{Gaiotto:2008sd}, such vacua were shown to arise naturally when D3 branes intersects
with D5 branes. In addition, we have found all supersymmetric Janus field theories where the coupling constant
$e^2$ depends on other spatial coordinates. 

In this work we repeat a similar analysis for the case where $\tau$ depends on spatial coordinates. 
In Refs.~\cite{Gaiotto:2008sd,Gaiotto:2008ak}, the SL(2,Z) transformation and the brane picture were an important tool to explore Janus-type field theories.    Especially 
the vacuum structure was explored in the detail. However, one can still ask whether there is nontrivial classical vacua besides the usual Coulomb vacua in our case. Our analysis shows that if such vacua exist, they would break the supersymmetry further to only two. However, the generalization of the Nahm equation is too
complicated at this moment. The BPS objects are dyonic objects, and their characterization is equally or more complicated than the previous case.  In the presence of the $\theta$ term, dyons would carry additional 
electric charge due to Witten effect~\cite{Witten:1979ey}.

One interesting simplification is a sharp interface where the coupling constant $\tau$ jumps from one constant
value to another in a very small region. As the $\theta$ angle jumps, such interface can be interpreted partially
as an axionic domain wall~\cite{Sikivie:1984yz,Lee:1986mm}. Electromagnetic wave reflected or transmitted through such a wall would have rotated polarization. We calculate the reflection and transmission coefficient.  We fully investigate the 1/2 BPS dyonic object near the wall, ignoring the non-Abelian core.

A full classification of four-dimensional Janus gauge  theory with partially conserved supersymmetry 
with spatial dependent coupling $e^2(x,y,z)$ has been done in our previous paper \cite{Kim:2008dj}.  We think that the same classification works also 
for the present case with $\tau(x,y,z)$, and have worked out all cases in detail.

One can introduce the interface degrees of freedom on a sharp interface without further breaking of supersymmetry as in Ref.~\cite{Gaiotto:2008sd}. Our result suggests that one could introduce a more general class of interface Lagrangian to our more general Lagrangian with more parameters and
less supersymmetries. It would be interesting to explore these Lagrangians and their properties.

This paper is presented as follows: In Sec.2, we review the 8
supersymmetric  Janus Yang-Mills theories in four-dimimensions. In Sec.3, we study the
vacuum structure of this theory. We raise the possibility of vacua preserving
only two supersymmetries. In Sec.4, we consider  1/2 BPS field configurations.
 In Sec.5, we focus on the sharp interface for the  coupling constant. The image charges
for the magnetic monopoles and electric charges are found. The
wave propagation and reflection at the interface is studied. In
Sec.6, less supersymmetric Janus Yang-Mills theories  whose coupling constants 
may have  additional space-time dependence are explored briefly.
 
(While this work is written, a   paper~\cite{Chen:2008tu} has appeared where there is  some overlap.
We feel some of points raised here seem new.)

\section{A Brief Review of 8 Supersymmetric Janus Lagrangian}

We start with the ten-dimensional supersymmetric Yang-Mills Lagrangian  is
\be {\cal L}_0 =- \frac{1}{4e^2} \Tr \Big(  F^{MN}F_{MN} +
 2i\bar{\Psi}\Gamma^M D_M \Psi  \Big) \ ,
\ee
where $M,N=0,1,2...,9$ and $\mu,\nu=0,1,2,3$. We use the ten-dimensional
notation for convenience. The gamma matrices $\Gamma^M$ are in the
Majorana representation, and the gaugino field $\Psi$ is Majorana
and Weyl.   The spatial signature is $(-+++...+)$. The Lagrangian
is invariant under the  supersymmetric transformation
\be \delta_0 A_M=i\bar{\Psi}\Gamma_M\epsilon\, , \qquad \delta_0\Psi =
\frac{1}{2} \Gamma^{MN}\epsilon F_{MN} \ , \label{susy0}  \ee
where the supersymmetry (SUSY) parameter $\epsilon$ is Majorana and satisfies the
Weyl condition,
\be   \Gamma^{012\cdots9}\epsilon =\epsilon \  . \label{weyl} \ee
As we consider $1+3$ dimensional space-time $x^\mu= x^0,x^1,x^2,x^3$, the
remaining spatial gradient $\partial_M=0$ with $M=4,5\cdots 9$ and
the gauge field $ A_M $ become scalar fields $\phi_M$ with
$M=4,5\cdots 9$. In four-dimensional space-time, one can have an
additional term in the Lagrangian
\ba {\cal L}_\theta =  -\frac{\theta}{32\pi^2}\Tr
\tilde{F}^{\mu\nu} F_{\mu\nu}  =  \frac{1}{8\pi^2} W^\mu \partial_\mu \theta\ , \ea
where the dual  field strength is $\tilde{F}^{\mu \nu}=
\epsilon^{\mu\nu\rho\sigma}F_{\rho\sigma}/2$ with
$\epsilon^{0123}=1$ and $W_\mu$ is the Chern-Simons term,
$W^\mu=\epsilon^{\mu\nu\nu\rho} 
\Tr(A_\nu \partial_\rho A_\sigma/2 -i A_\mu A_\nu A_\rho/3)$.
As ${\cal L}_\theta$ is  a total derivative,
the supersymmetry of the original Lagrangian ${\cal L}_0$ would be
intact.

We are interested in the case where the coupling constants $e^2,
\theta$  depend on space-time coordinates.  The original
Lagrangian ${\cal L}_0+{\cal L}_\theta$ is no longer invariant
under the original  supersymmetric transformation modulo a total
space-time derivative. Fortunately, one can maintain some of supersymmetries if 
one modifies the supersymmetric transformation of the gaugino field by
$\delta_1\Psi$ and also the Lagrangian by   additional terms,
which depend on the  derivatives of the coupling constant. 
For this work,  we need to take the space-time dependent  
supersymmetric parameter  $\epsilon(x)$. The Lagrangian 
${\cal L}_0+{\cal L}_\theta$ transforms
under  the   supersymmetric transformation $\delta_0$ of
Eq.~(\ref{susy0}) nontrivially as follows:
\ba && \delta_0 ({ \cal L}_0 +{\cal L}_\theta )= -
\partial_\mu\Big(  \frac{1}{4e^2}\Big) \Tr \left( F_{MN} \ i
\bar{\Psi}\Gamma^{MN}\Gamma^\mu\epsilon  \right)
\nonumber \\
&& \ \ \ \   -\frac{1}{2e^2} \Tr ( F_{MN}\,
i\bar{\Psi}\Gamma^\mu\Gamma^{MN}\partial_\mu\epsilon) +
 \Big(\frac{\partial_\mu \theta}{16\pi^2}\Big) \epsilon^{\mu\nu\rho\sigma} \Tr (
F_{\nu\rho}\ i\bar{\Psi}\Gamma_\sigma\epsilon ) \ . \label{d0l0} \ea
The additional transformation of the original Lagrangian due to
$\delta_1 \Psi$ would be
\be \delta_1{\cal L}_0 = -\partial_\mu \left(\frac{1}{2e^2}\right)
i\bar\Tr ( {\Psi}\Gamma^\mu \delta_1 \Psi )
-\frac{1}{e^2} \Tr( i\bar{\Psi}\Gamma^M D_M\delta_1 \Psi )\ . \label{d1l0}  \ee

Let us focus on the case where the coupling constants  $e^2,
\theta $ depends only on the $x^3=z$ coordinate.
It has been shown recently in Ref.~\cite{Gaiotto:2008sd} that the
half of the original supersymmetry could be maintained if the
spatial dependence of two coupling constants is  constrained so
that
\be\frac{1}{e^2} = D\sin2\psi\ , \qquad  \theta =  \theta_0   + 8\pi^2
D \cos2\psi \ . \label{psid} \ee
with the space-time dependence arising only from $\psi(z)$, which can be an 
arbitrary function. 
 Note that in the limit $D\rightarrow \infty$, 
$\theta_0 \rightarrow \mp\infty$ and $\psi(z) \rightarrow 0,\pi/2$ with
the combinations $D\psi(z)$ and $\theta_0 \pm 8 \pi^2 D$ kept finite,
the space-time dependence appears only in the
fine structure constant $4\pi/e^2$. 
Notice also that a constant shift
of $\theta$ by $2\pi$ does not change physics. The complex
coupling constant becomes
\be \tau = \frac{\theta}{2\pi}+ \frac{4\pi i}{e^2} =\tau_0 + 4\pi
D e^{2i \psi},  \ee
where $\tau_0=\theta_0/(2\pi)$. 

With the coupling constants given by Eq.(\ref{psid}),   eight of the original 16
supersymmetries can be preserved~\cite{Gaiotto:2008sd}. The condition on
the supersymmetric parameter $\epsilon$ compatible with the Weyl condition
(\ref{weyl}) is
\be
 \epsilon(z)=e^{-\frac{\psi(z)}{2} \Gamma^{0123}}\epsilon_0  \label{susy}\ee
with a constant spinor $\epsilon_0$ such that
\be \Gamma^{3456}\epsilon_0=\epsilon_0 \ . \label{condition1}\ee
This condition also breaks the
global $SO(6)$ symmetry which rotates $4,5,6,7,8,9$ indices to
$SO(3)\times SO(3)$, each of which
 rotates $4,5,6$ and $7,8,9$ indices respectively.
As $\Gamma^{012...9}\epsilon=\epsilon$, we get
$\Gamma^{3456}e^{\psi\Gamma^{0123}}\epsilon=\epsilon$. 
The condition on $\epsilon_0$ is identical to the case with the constant $\theta$. 

As the $SO(6)$ symmetry is broken to $SO(3)\times SO(3)$, 
we split six scalar fields to
two sets each of which are made of three scalar fields. We denote
\be X_a= (X_1,X_2,X_3)= (\phi_4,\phi_5,\phi_6)\ , \ \ 
Y_a =(Y_1,Y_2, Y_3)=(\phi_7,\phi_8,\phi_9)\ . \ee
We will also interchangeably use
$ (X_1,X_2,X_3)= (X_4,X_5,X_6)$ and $(Y_1,Y_2, Y_3)=(Y_7,Y_8, Y_9)$. 
The indices for $\Gamma^a$ follow the indices for the scalar field
whenever they are contracted.
To cancel some of the terms in the zeroth order variation of the
original Lagrangian (\ref{d0l0}), one needs to add a correction to
the SUSY transformation of the gaugino field and also corrections
to the original Lagrangian. The correction to the original SUSY 
transformation (\ref{susy0}) is
\be \delta_1 A_M=0\; , \;\;  \delta_1\Psi =\psi'\Gamma^3( 
(\Gamma\cdot X) \cot\psi -  (\Gamma\cdot Y) \tan\psi ) \epsilon\  ,
   \ee
where the prime means $d/dz$. The correction to the original
Lagrangian is made of two parts. The first correction, which
depends on the first order in the derivative of the couple
constant, is given as
\ba {\cal L}_1 &=&  \frac{\psi'}{4e^2}\Tr \,    i\bar{\Psi}
\big( - \Gamma^{012}   +\frac{1}{\sin\psi} \Gamma^{456}
-\frac{1}{\cos\psi} \Gamma^{789}\big) \Psi 
\nonumber \\
& &
 + \frac{2\psi'}{e^2}\Tr \Big( -\frac{i}{\sin\psi} X_1[X_2,X_3] +\frac{i}{\cos\psi} Y_1[Y_2,Y_3] \Big)
 \ .
\ea
The second correction is quadratic in the derivatives of $\psi$ so
that
\ba && {\cal L}_2=-\frac{1}{2e^2}\Tr\left[ (
\psi'^2-(\psi'\cot\psi)')X_aX_a +(\psi'^2 +(\psi'\tan\psi)')Y_aY_a
\right] \ .  \nonumber \\ && \ea
The total Lagrangian 
${\cal L}= {\cal L}_0+{\cal L}_\theta + {\cal L}_1+{\cal L }_2$ 
is invariant under the corrected SUSY transformation,
\ba  \delta A_M &=& (\delta_0 +\delta_1) A_M
  = i\bar{\Psi}\Gamma_M\epsilon, \nonumber \\
\delta \Psi &=& (\delta_0 +\delta_1)\Psi \nonumber \\
& = &
\frac{1}{2}F_{MN}\Gamma^{MN}\epsilon 
+ \psi'\Gamma^3 (\cot\psi (\Gamma\cdot X) -\tan\psi (\Gamma\cdot Y)) \epsilon \ .
\label{susy2} \ea

Redefine the scalar fields so that
\be \tilde{X}_a = X_a\sin\psi\  , \ \ \tilde{Y}_a = Y_a\cos\psi \ . \ee
The correction to the   supersymmetric transformation of the gaugino field can be absorbed as
\be \delta \Psi =\Big( \frac12 F_{\mu\nu}\Gamma^{\mu\nu} +\frac{1}{\sin\psi } \Gamma^{\mu} D_\mu\tilde{X}\cdot \Gamma+\frac{1}{\cos\psi}\Gamma^\mu D_\mu \tilde{Y}\cdot\Gamma+\cdots\Big)\  \epsilon .  \ee
The ${\cal L}_2$ can be absorbed into the scalar kinetic energy
\be {\cal L}_0+{\cal L}_2=...-\frac{1}{2e^2} \Big( \frac{1}{\sin^2\psi} (D_\mu \tilde{X}^a)^2 +\frac{1}{\cos^2\psi}
(D_\mu \tilde{Y}^a)^2\Big)+...\ . \ee
The whole  Lagrangian ${\cal L}$ becomes
\ba {\cal L} &=& - \frac{1}{4e^2} \Tr \Big( F^{\mu\nu}F_{\mu\nu} 
+ \frac{e^2\theta}{8\pi^2} \tilde{F}^{\mu\nu} F_{\mu\nu}
+ \frac{2}{\sin^2\psi} D^\mu \tilde{X}_a D^\mu \tilde{X}^a 
+ \frac{2}{\cos^2\psi} D_\mu \tilde{Y}^aD^\mu \tilde{Y}^a    \Big)
 \nonumber \\
& &+ \frac{1}{4e^2}\Tr\Big( \frac{1}{\sin^4\psi} [\tilde{X}^a,\tilde{X}^b]^2 +\frac{1}{\cos^4\psi} [\tilde{Y}^a,\tilde{Y}^b]^2+ 
\frac{8}{ \sin^22\psi} [ \tilde{X}^a,\tilde{Y}^b]^2\Big) \nonumber
\\ & &
-\frac{1}{2e^2}\Tr \Big( i \bar{\Psi}\Gamma^\mu D_\mu \Psi
+\frac{1}{ \sin\psi} \bar{\Psi}\Gamma^a[\tilde{X}^a,\Psi] +\frac{1}{\cos\psi} 
\bar{\Psi}\Gamma^p[\tilde{Y}^p,\Psi] \Big) \nonumber \\
& &  +  \frac{\psi'}{4e^2}\Tr \,    i\bar{\Psi}
\big( - \Gamma^{012}   +\frac{1}{\sin\psi} \Gamma^{456}
-\frac{1}{\cos\psi} \Gamma^{789}\big) \Psi
\nonumber \\
& &
 + \frac{2\psi'}{e^2}\Tr \Big( -\frac{i}{\sin^4\psi} \tilde{X}_1[\tilde{X}_2,\tilde{X}_3]  
 +\frac{i}{\cos^4\psi} \tilde{Y}_1[\tilde{Y}_2,\tilde{Y}_3] \Big) \ . \label{lagr}\ea
The combined SUSY transformation (\ref{susy2}) becomes
\ba && \delta A_\mu = i\bar{\Psi} \Gamma_\mu \epsilon, \qquad
\delta \tilde{X}_a = \frac{1}{\sin\psi}\bar{\Psi}\Gamma_a
\epsilon,\qquad
\delta \tilde Y_p = \frac{1}{\cos\psi} \bar{\Psi}\Gamma_p \epsilon \nonumber \\
&& \delta \Psi = \Big( \frac{1}{2}F_{\mu\nu}\Gamma^{\mu\nu} +\frac{1}{\sin\psi}
D_\mu\tilde{X}_a\Gamma^{\mu a} + \frac{1}{\cos\psi}D_\mu\tilde{Y}_p \Gamma^{\mu p }
-\frac{i}{\cos\psi\sin\psi} [\tilde{X}_a,\tilde{Y}_p]\Gamma^{ap}\nonumber
\\
&& \qquad
 -\frac{i}{2\sin^2\psi} [\tilde{X}_a,\tilde{X}_b] \Gamma^{ab}
-\frac{i}{2\cos^2\psi}[\tilde{Y}_p,\tilde{Y}_q] \Gamma^{pq}\Big) \epsilon \ . \label{susy3} \ea
We can choose the gauge group to be any simple Lie group $G$.

\section{The Vacuum Structure}

The energy density from the above Lagrangian (\ref{lagr}) is not positive definite. To consider the vacuum, we
put $A_\mu=0$ and $\Psi=0$. We also put the scalar fields can depend only on $x^3=z$ coordinates. 
Only nontrivial part of the energy density is
\ba {\cal E} &=& \frac{1}{2e^2} \Tr  \Big( \frac{  \tilde{X}_a'^2}{\sin^2\psi} + \frac{  \tilde{Y}_a'^2}{\cos^2\psi} 
-\frac{[\tilde{X}_a,\tilde{X}_b]^2}{2\sin^4\psi} -\frac{[\tilde{Y}_a,\tilde{Y}_b]^2}{2\cos^4\psi} -\frac{[\tilde{X}_a, \tilde{Y}_b]^2 }{\cos^2\psi\sin^2\psi }\Big)  \nn \\
& & +\frac{2i\psi'}{e^2} \Tr \Big( \frac{1}{\sin^4\psi} \tilde{X}_1[\tilde{X}_2,\tilde{X}_3] - \frac{1}{\cos^4\psi}
\tilde{Y}_1[\tilde{Y}_2,\tilde{Y}_3] \Big) \ . \ea
We rewrite the above energy density as
\ba {\cal E} &=& \frac{1}{2e^2}\Tr  \Big( \frac{  \tilde{X}'_a}{\sin\psi} + \frac{i}{2} \epsilon^{abc}([X_b,X_c]-[Y_b,Y_c])\cos\psi -i  \epsilon^{abc}[X_b,Y_c] \sin\psi \Big)^2 \nn \\
& &+ \frac{1}{2e^2} \Tr \Big( \frac{ \tilde{Y}_a'}{\cos\psi}  -\frac{i}{2}\epsilon^{abc}([X_b,X_c]-[Y_b,Y_c])\sin\psi-i\epsilon^{abc}[X_b,Y_c]\cos\psi\Big)^2 \nn \\
& & + \frac{1}{2e^2} \Big(\sum_a  i [X_a,Y_a] \Big)^2 + {\cal E}_b \ . 
\ea
The boundary term is
\ba 
{\cal E}_b &=& -2iD\, \Tr \Big(  \tilde{X}_1[\tilde{X}_2,\tilde{X}_3]\cot^2\psi+\tilde{Y}_1[\tilde{Y}_2,\tilde{Y}_3] 
\tan^2\psi \Big)'  \nn \\
& & + iD\epsilon^{abc}\Tr\Big(\tilde{X}_a[\tilde{Y}_b,\tilde{Y}_c]+\tilde{Y}_a[\tilde{X}_b,\tilde{X}_c]\Big)' \ .  \ea
 Assuming the boundary contributions vanish, the energy would be non-negative and the classical vacuum would
 satisfy the vacuum equations
 \ba && \frac{   \tilde{X}'_a}{\sin\psi} + \frac{i}{2} \epsilon^{abc}([X_b,X_c]-[Y_b,Y_c])\cos\psi -i  \epsilon^{abc}[X_b,Y_c] \sin\psi  =0\,  , 
 \nn \\
 && \frac{ \tilde{Y}'_a}{\cos\psi}  -\frac{i}{2}\epsilon^{abc}([X_b,X_c]-[Y_b,Y_c])\sin\psi-i\epsilon^{abc}[X_b,Y_c]\cos\psi=0\,  . \label{vaceq}  \ea
Of course the obvious vacuum configurations satisfying the above equations are the Abelian Coulomb vacua, 
\be \tilde{X}_a'=0,\qquad 
\tilde{Y}_a'=0, \qquad
[ X_a,X_b]=[Y_a,Y_b]=[X_a,Y_b]=0 . \ee
Thus,   $\tilde{X}_a , \tilde{Y}_a$ are constant and can be diagonalized. As $\delta \Psi=0$ of (\ref{susy3}), these Abelian vacua are fully supersymmetric. 

The interesting question is whether there exist any nontrivial solution for the vacuum Eq.~(\ref{vaceq}).
For a constant $\theta$ case, this vacuum equation turns out to be the Nahm equation for the magnetic monopoles, and nontrivial vacuum are allowed when $e^2$ vanishes on some planes. This has nice
interpretation as D3 branes intersecting with D5 branes. In the present case, it is not clear at all whether nontrivial,
or non-Abelian, solutions exist.  If they do,
one may wonder the number of supersymmetries preserved. Let us consider the generic case.  The SUSY transformation for the vacuum configuration would be 
\ba \delta \Psi &=& \Big( \frac{1}{\sin\psi} \tilde{X}_a' \Gamma^{3a} +\frac{1}{\cos\psi}\tilde{Y}_p'\Gamma^{3p} 
\nonumber \\
& & -
\frac{i}{2} [X_a,X_b]\Gamma^{ab} -\frac{i}{2}[Y_p,Y_q]\Gamma^{pq} 
-i [X_a,Y_p]\Gamma^{ap} \Big)\epsilon \, .  \ea
It vanishes for the vacua satisfying the vacuum Eq.~(\ref{vaceq}) only if  
 the SUSY parameter $\epsilon_0$ satisfies the additional conditions
\be \Gamma^{3489}\epsilon_0=\Gamma^{3597}\epsilon_0=\Gamma^{3678}\epsilon_0=-\epsilon_0 \, .
\label{navac} \ee
As only two of the above conditions are independent,  any  generic non-Abelian vacuum, if exist,  would break
the number of supersymmetries to two.

When $\theta$ is constant, it is known that there may be nontrivial vacua when the $e^2(z)$ vanishes
in some points\cite{Kim:2008dj}.  When D3 branes are connecting D5 branes, the dilation field vanishes at the location of D5 branes and so the coupling constant $e^2(z)$ in the theory on D3 branes varies while vanishing at the D5 locations. The vacua of the theory characterizes D3 branes. After T-dual transformation to D1-D3, the vacuum structure characterizes how D1 branes end on D3 branes. As D1 branes ending on D3 branes  appear as magnetic monopoles, we know that the Nahm equation characterizes D1 branes ending on D3 branes.  
The S-dual version of the above vacuum configuration, which appears as the boundary field theory, has been studied by  Gaiotto and Witten \cite{Gaiotto:2008ak}.    

\section{ BPS equations}

Similar to the constant  $\theta$ case, we can introduce 
two possible 1/2 BPS conditions:
\be \Gamma^{1234}\epsilon_0= \epsilon_0 \, , \qquad  
\Gamma^{70}\epsilon_0= \epsilon_0\, .   \label{12cond} \ee
The first one was for magnetic monopoles and the second one was for the charged
massive particles in the Coulomb phase.  
With nontrivial $\theta$, the  both conditions become  
the 1/2 BPS condition  for   dyons.
If we impose both conditions, we would get 1/4 BPS configurations. If there is nontrivial
non-Abelian vacuum of only two supersymmetries (\ref{navac}), the above BPS condition would be
incompatible, implying that there would be no BPS dyons in such a  non-Abelian vacuum.

With the first condition of (\ref{12cond}) in the Coulomb vacuum,  
one can read the 1/2 BPS equations for dyons from the supersymmetric 
transformation to be  $Y_a=0$.
The remaining 1/2 BPS equations is
\ba  \label{12bps}
&& E_i - D_i \tilde X_1 = 0, \nonumber\\
&& (B_1+iB_2)-\cot\psi (D_1+iD_2)\tilde{X}_1= 0 ,  \nonumber\\
&& B_3 - \cot\psi D_3 \tilde X_1
        - \frac{i}{\sin^2\psi} [\tilde X_2, \tilde X_3] = 0, \nonumber \\
&&   
(D_1+iD_2) (\tilde X_2 + i \tilde X_3)  = 0, \nonumber \\
&& D_3(\tilde X_2 + i \tilde X_3)
       - \cot\psi [\tilde X_1, \tilde X_2 + i \tilde X_3] = 0. \ea
where the electric and magnetic components of the field strength are
\be E_i=  F_{i0}\, , \qquad  B_i = \frac12 \epsilon_{ijk} F_{jk}\, . \ee
These equations are consistent with the Gauss law
\be \label{gauss}
D_i \Big(\frac{1}{e^2}E_i +\frac{\theta}{8\pi^2} B_i \Big) 
+ \frac{i}{e^2} \Big( [X_a, D_0 X_a] + [Y_a,D_0Y_a]\Big)= 0 \, . \ee
One could impose a further constraint
$\tilde{X}_2=\tilde{X}_3= 0$, and then the above BPS equations become
\be 
B_i -\cot\psi D_i \tilde{X}_1=0\, , \qquad
E_i - D_i \tilde X_1 = 0\, .
\label{halfbps} \ee

These equations can also be obtained from the energy functional, 
which can be reshuffled to
\begin{align}
{\cal H} &= \int d^3x \frac1{2e^2} \Tr \left[ (E_i-D_i\tilde X_1)^2 
 + (B_1 - \cot\psi D_1 \tilde X_1)^2 \right. \nonumber \\
 &+ (B_2 - \cot\psi D_2 \tilde X_1)^2 
 + \left(B_3 - \cot\psi D_3 \tilde X_1 
      - \frac{i}{\sin^2\psi}[\tilde X_2, \tilde X_3]\right)^2  \nonumber \\
 &+ \frac1{\sin^4\psi}|D_1(\tilde X_2 +i \tilde X_3)
                        + iD_2(\tilde X_2 +i \tilde X_3) |^2 \nonumber \\
 &+ \left( \frac i{\sin^2\psi} [\tilde X_2, \tilde X_3]
           - \cot\psi D_3\tilde X_2 + D_0 \tilde X_3 \right)^2 \nonumber\\
 &+ \left. \left( \frac i{\sin^2\psi} [\tilde X_2, \tilde X_3]
           + \cot\psi D_3\tilde X_3 + D_0 \tilde X_2 \right)^2 \right]
     \nonumber \\
 &+ \int \partial_i \Tr \frac1{e^2} \tilde X_1 \left(
    E_i - \tan\psi B_i
    + \delta_{i3} \frac{i}{\sin^2\psi}[\tilde X_2,\tilde X_3] \right)
  - \left. \frac{2D}{\sin^2\psi} \Tr \tilde X_1 [\tilde X_2,\tilde X_3] 
    \right|_{z=-\infty}^{z=\infty} \nonumber \\
 &+ 2D \int \partial_i \Tr B_i \tilde X_1  + \text{($Y$-dependent terms)}\, ,
\end{align}
where we have used the Gauss law in completing the squares.
Note that there are three boundary terms. Among these, the first term
vanishes on imposing the BPS equations and the second term is zero for Coulomb
vacua. Therefore, for the half-BPS configurations the energy is proportional to
the magnetic charge
\be \label{12energy}
{\cal H} = 2D \int d^3x \partial_i \Tr B_i \tilde X_1\, .
\ee
Thus the 1/2 BPS configuration with nonzero energy would be those with $X_2=X_3=0$ and
satisfies Eq~.(\ref{halfbps}).

The 1/2-BPS equations for the supersymmetric condition 
$\Gamma^{70} \epsilon_0 = \epsilon_0 $ can be obtained in a similar manner.
The resulting equations and the energy are exactly the same as 
Eqs.(\ref{12bps}) and (\ref{12energy}) with $\tilde X_a$ and $\psi$ replaced by 
$\tilde Y_a$ and $\pi/2 - \psi$, respectively. After imposing both conditions, 
we got the 1/4 BPS equations, which are complicated and mixed version of the above 1/2 BPS 
equations. One novel aspect of 1/4 BPS equations is that the electric and magnetic fields
are not parallel to each other.

Since the solution of the BPS equation is a dyonic object in the presence
of $\theta$ term, we would like to briefly discuss the Witten effect 
\cite{Witten:1979ey} in this case. Assume that the vacuum is given by
\be
\Tr \phi \phi = \tilde v^2,
\ee
where $\phi = \tilde X_1$ or $\phi = \tilde Y_1$ depending of
the supersymmetric condition.
The Noether charge $n$ generating the gauge transformation around the 
direction $\phi$ is 
\begin{align}
n &= \int d^3x \frac{\partial {\cal L}}{\partial(\partial_0 A_i^a)} \delta A_i^a
           \nonumber \\
 &= \int d^3x \Tr \left( \frac1{e^2}E_i + \frac\theta{8\pi^2}B_i \right)
                  \frac1{\tilde v}D_i \phi\, .
\end{align}
This is quantized as an integer.
Using Eqs.(\ref{psid}) and (\ref{halfbps}), we find
\ba  \label{qm}
n &=& \left( \frac{\theta_0}{8\pi^2} \pm D \right) \int d^3x 
             \partial_i \Tr (B_i \phi/\tilde v ) \nonumber \\
   &=& \left( \frac{\theta_0}{8\pi^2} \pm D \right) Q_M \nonumber \\
   &=& ( \tau_0 \pm 4\pi D ) n_m\, ,
\ea
where the upper (lower) sign is for $\phi = \tilde X_1\ (\tilde Y_1)$.
$Q_M$ is the magnetic charge quantized as $Q_M = 4\pi n_m$ 
with $n_m$ being an integer. Therefore, for 1/2-BPS solutions to survive
quantum mechanically, the parameters in the coupling should be quantized.
Note, however, that there is no simple relation
between the electric charge and the magnetic charge if the couplings
are not constants. For example, with Eq.(\ref{halfbps}),
\be  \label{qe}
Q_E = \int d^3x \partial_i \Tr 
            \left( \frac1{e^2 \tilde v} E_i \tilde X_1 \right)
    = 2D \int d^3x \partial_i \Tr 
            \left( \frac{\sin^2\psi}{\tilde v} B_i \tilde X_1 \right).
\ee
The electric charge would have been a sum of Noether charge and that of the Witten effect
if $\theta$ is constant. 

\section{Sharp Interface}
Here, we consider the 1/2-BPS case that the coupling constants $e^2(z), \theta(z)$ 
change from one value to another at a sharp interface so that
\be
(e, \theta,\psi ) = \begin{cases} (e_1, \theta_1, \psi_1) & \text{for }z>0 \\
                            (e_2, \theta_2,\psi_2) & \text{for }z<0 
              \end{cases} .
\ee
As there is no matter source at the interface, 
we get various continuity conditions from the equations of motion.
The following quantities are continuous at the interface $z=0$:
\begin{align} \label{bdry}
&E_i,\ \left( \frac1{e^2}E_3 + \frac\theta{8\pi^2} B_3 \right),\ 
B_3,\ \left( \frac1{e^2}B_i - \frac\theta{8\pi^2} E_i \right),\nonumber\\
&\tilde X_a,\ D_i \tilde X_a ,\ \ \cot\psi D_3 \tilde X_a\, , \ 
\tilde Y_p,\ D_i \tilde Y_p ,\ \tan\psi D_3 \tilde Y_p\, ,  \qquad (i=1,2).
\end{align}
{}From this boundary condition we see that, if $\theta$ is not constant, 
electric charge is induced on the boundary, which is proportional to the 
magnetic flux $\Phi$ through the boundary
\be \label{indu}
Q_E^\textrm{induced} = \int \partial_i \left( \frac1{e^2} E_i \right)
= \frac{\theta_2 - \theta_1}{8\pi^2} \Phi\, .
\ee

\subsection{Reflection and transmission of waves}
Let us consider now a massless wave propagating toward the interface
of the two coupling constants from $z>0$ region. 
The fields and their derivatives in (\ref{bdry}) should be continuous
across the interface $z=0$. 
A part of the incident wave will be reflected and the 
rest may get refracted or transmitted. Let us
call the electromagnetic field of the incident wave to be ${\bf
E},{\bf B}$, the reflected wave to be ${\bf E}'', {\bf B}''$ and
the transmitted wave to be ${\bf E}',{\bf B}'$. The continuity equations at $z=0$ are
\ba && \big({\bf E}+ {\bf E}''-{\bf E}'\big) \times \hat{\bf z}= 0 \, , 
\nonumber
\\
&& \big( {\bf B}+{\bf B}''-{\bf B} \big) \cdot \hat{\bf z} = 0\, , 
\nonumber \\
&& \bigg( \frac{ {\bf E}+{\bf E}''}{e_1^2} -\frac{{\bf E}'}{e_2^2} 
 + \frac{\theta_1}{8\pi^2}({\bf B} + {\bf B}'')
 - \frac{\theta_2}{8\pi^2} {\bf B}' 
\bigg) \cdot \hat{\bf z}= 0 \, , \nonumber \\
&& \bigg(\frac{{\bf B}+{\bf B}''}{e_1^2} -\frac{{\bf B}'}{e^2_2}
 - \frac{\theta_1}{8\pi^2}({\bf E} + {\bf E}'')
 + \frac{\theta_2}{8\pi^2} {\bf E}' 
\bigg) \times \hat{\bf z} = 0\, .  \label{conti2} \ea
The space-time dependence waves would be  $e^{-iwt+i{\bf k}\cdot
{\bf x}}$, $e^{-iwt+i{\bf k}''\cdot {\bf x}}$, and 
$e^{-iwt+i{\bf k}'\cdot {\bf x}}$ for the incident, reflected, and transmitted
waves, respectively. The wave equation at each region and the
above continuity equations imply that
\be w= |{\bf k}|=|{\bf k}''|=|{\bf k}'|, \;\; {\bf k}={\bf k}' ,
\;\;  ({\bf k}+{\bf k}'')\cdot \hat{z}=0\, .  \ee
Thus the transmitted wave is not refracted at all. After taking
out the space-time dependence,  we can express the electric fields
of the reflected and transmitted waves in terms of the the incident wave. 
The amplitudes of the transmitted electric fields is given by
\be
{\bf E}'_0 = \frac{\sin2\psi_1}{\sin(\psi_1+\psi_2)}
 [\cos(\psi_1-\psi_2) {\bf E}_0 - \sin(\psi_1-\psi_2) {\bf B}_0 ] \, .
\ee
Note that the polarization direction is rotated. For the reflected
electric field, the expression explicitly depends on the incident angle
and will not be shown here since it is rather complicated.
However, if the incident electric field has the form
${\bf E}_0 = E_0(\cos2\psi_1 \hat{\bf m}+ \sin2\psi_1 \hat{\bf n})$,
where $\hat{\bf m}$ is the unit vector normal to the plane formed
by $\hat{\bf z}$ and ${\bf k}$ and 
$\hat{\bf n} = \hat{\bf k} \times \hat{\bf m}$,
it can be written in a simple form
\ba
{\bf E}''_0 &=& E_0 \frac{\sin(\psi_1-\psi_2)}{\sin(\psi_1+\psi_2)}
     \hat{\bf m} \nonumber \\
    &=& \frac{\sin(\psi_1-\psi_2)}{\sin(\psi_1+\psi_2)}
 (-\cos2\psi_1 {\bf E}_0 + \sin2\psi_1 {\bf B}_0 ).
\ea
The expression in the second line also holds when 
the incident wave is normal to the $xy$ plane.
The corresponding magnetic fields can be obtained from the relation
${\bf B} = \frac{{\bf k}}{w} \times {\bf E}$.
The reflection and the transmission coefficients defined as
$E_0'' = r E_0, E_0'= t E_0$ are however independent of the details of
the incident wave and are always given by
\be r = \left|\frac{\sin (\psi_1-\psi_2)}{\sin(\psi_1+\psi_2)}\right|,\qquad
 t = \left|\frac{\sin 2\psi_1}{\sin(\psi_1+\psi_2)}\right|. \ee

\subsection{Gauge fields of a single dyon in the Abelian limit}

For simplicity, we consider the SU(2) gauge theory which is broken spontaneous
to U(1) subgroup by the Higgs expectation values at the vacuum
\be
\langle \tilde X_1 \rangle =  \frac{\sigma_3}{\sqrt2} \tilde v.
\ee
The diagonal components of the fields will be massless, and off-diagonal ones
will be massive. We solve the 1/2 BPS Eq.~(\ref{halfbps})
in the Abelian limit where the non-Abelian core size vanishes. 
For definiteness, we work with the first condition of Eq.(\ref{12cond}).
For a single dyon with charge $(q,g)$ at $z=a>0$, electric and
magnetic fields would have the form
\begin{align}
\mathbf{B} &= \begin{cases}
\frac{g}{4\pi} \frac{(x,y,z-a)}{r_+^3}
     + \frac{g'}{4\pi} \frac{(x,y,z+a)}{r_-^3}, &z>0 \\
\frac{g''}{4\pi} \frac{(x,y,z-a)}{r_+^3}, &z<0 \ ,
\end{cases} \nonumber \\
\mathbf{E} &= \begin{cases}
\frac{e_1^2q}{4\pi} \frac{(x,y,z-a)}{r_+^3}
     + \frac{e_1^2q'}{4\pi} \frac{(x,y,z+a)}{r_-^3}, &z>0 \\
\frac{e_2^2q''}{4\pi} \frac{(x,y,z-a)}{r_+^3}, &z<0 \ .
\end{cases}
\end{align}
where $r_\pm^2 = x^2 + y^2 + (z\mp a)^2$ and the group factor $\sigma_3/\sqrt2$
is omitted for simplicity. The image charges $(q',g')$ and $(q'',g'')$ are
to be determined from the BPS equations and the boundary conditions.
The configuration of scalar field $\tilde X_1$
may be obtained by integrating the electric field through 
$E_i = D_i \tilde X_1$. From the BPS equation $E_i = \tan\psi B_i$,
it immediately follows that
\be \label{qgrel}
q = \frac{g}{e_1^2} \tan\psi_1, \qquad
q' = \frac{g'}{e_1^2} \tan\psi_1,\qquad
q'' = \frac{g''}{e_2^2} \tan\psi_2.
\ee
In addition, we have four equations from the boundary conditions 
Eq.(\ref{bdry}). Since we have only three unknown charges it may look
over constrained. However, with the help of the relation Eq.(\ref{psid})
between $e^2$ and $\theta$, we can find a solution satisfying all the
equations,
\ba
g'  &=& -g \frac{\sin(\psi_1-\psi_2)}{\sin(\psi_1+\psi_2)} \nonumber \\
g'' &=& 2g \frac{\sin\psi_1 \cos\psi_2}{\sin(\psi_1+\psi_2)} \nonumber \\
q' &=& -2gD \frac{\sin^2\psi_1 \sin(\psi_1-\psi_2)}{\sin(\psi_1+\psi_2)} 
                   \nonumber \\
q'' &=& 4gD \frac{\sin\psi_1 \sin^2\psi_2 \cos\psi_2}{\sin(\psi_1+\psi_2)} 
\ea
The magnetic and electric fluxes to the northern and southern hemispheres are,
respectively,
\ba
\Phi_M^N &=& \frac12 (g + g'), \qquad \Phi_M^S = \frac12 g'', \nonumber\\
\Phi_E^N &=& \frac12 (q + q'), \qquad \Phi_E^S = \frac12 q'', \nonumber\\
\ea
and the charges $Q_M$ and $Q_E$ in Eqs.(\ref{qm}) and (\ref{qe}) are
given by the total fluxes 
\ba
Q_M &=& \frac12(g + g' + g'') = g, \nonumber \\
Q_E &=& \frac12(q + q' + q'') 
          = 2gD\sin\psi_1 \sin\psi_2 \cos(\psi_1-\psi_2),
\ea
which satisfy Eq.(\ref{qe}) as it should be.

One may wonder whether this object has nonzero field angular momentum.
At a first look this has to be the case because a dyon produces
nonzero angular momentum in the background of an
axionic domain wall \cite{Lee:1986mm} for which $\theta$ changes from
zero to $2\pi$, while the coupling $e^2$ remains a constant. For the
present case, however, the electric field
is proportional to the magnetic field thanks to the half-BPS equation and
hence the field angular momentum defined by
\begin{equation}
\mathbf{M} = \int d^3x\, \mathbf{r} \times (\mathbf{E} \times \mathbf{B}) \, .
\end{equation}
is identically zero. 

The magnetic flux through the $xy$ plane is 
\begin{equation}
\Phi_M^{a>0} = -\Phi_M^S = -\frac12 g''\, ,
\end{equation}
which induces electric charge on the boundary as in Eq.(\ref{indu}).
If the dyon is on the negative $z$ axis ($a<0$), 
the corresponding magnetic flux on the $xy$ plane would be
\begin{equation}
\Phi_M^{a<0} = \frac12 g''|_{1\leftrightarrow 2}\, .
\end{equation}

Let us now consider the situation that
a dyon with charge $(\tilde q,g)$ at $z<0$ region passes from the
$xy$ plane to $z>0$, where $\tilde q = g \tan\psi_2/e_2^2$ as given
in (\ref{qgrel}). Since the coupling constants change from 
$(e_2, \theta_2)$ to $(e_1, \theta_1)$, the charge should change to 
$(q,g)$, accordingly. It is interesting to check how the conservation
of electric charge works. In fact, as the dyon passes the $xy$ plane
the induced electric charge (\ref{indu}) should also change due to the
change of magnetic flux, which is given by
\ba
\Delta\Phi &=& \Phi_M^{a>0}(q,g) - \Phi_M^{a<0}(\tilde q,g) \nonumber \\
&=& - g \frac{\sin\psi_1 \cos\psi_2}{\sin(\psi_1+\psi_2)} 
   - g \frac{\sin\psi_2 \cos\psi_1}{\sin(\psi_1+\psi_2)} \nonumber \\
&=& - g\, .
\ea
Then,
\be \label{qin}
\Delta Q_E^\textrm{induced} = \frac{\theta_1 - \theta_2}{8\pi^2} g\, .
\ee
This is to be compared with the change of the electric charge of the dyon
\begin{equation}
\Delta q = q - \tilde q = 
\frac{g}{e_1^2} \tan\psi_1 - \frac{g}{e_2^2} \tan\psi_2\, ,
\end{equation}
where we have used Eq.(\ref{qgrel}). On using Eq.(\ref{psid}), 
this is precisely cancelled by $ \Delta Q_E^\textrm{induced} $ in
Eq.(\ref{qin}). Usually the induced charge on the axion domain wall is due
to the polarization of fermions, which led the photon-axion interaction. In our 
case, a further study is needed to clarify the exact nature of the sharp interface.
 
If the 1/4-BPS configurations are considered, the electric and magnetic fields are not
parallel to each other in general, and more richer configurations with nonzero angular
momentum would appear near a sharp interface. The detail will be left as an exercise.

\section{Additional Susy Breaking Janus}

Let us consider the further supersymmetry breaking Janus
configurations.  This could happen two ways. First is to have additional terms
in the Lagrangian while keeping the coupling constant  $\tau(z)$ depending only on 
one spatial coordinate. Another is to introduce additional space-time dependence
to the coupling constant and then correct the Lagrangian.  We have done
a full classification in Ref.~\cite{Kim:2008dj} without the $\theta$ term,
 and  the same classification works
for the present case as long as we keep the supersymmetric condition on
$\epsilon_0$, which we describe in the following.
 
\subsection{the $\tau(z) $ case}

In this subsection we are still interested in the
case where the coupling constant $e^2(z),\theta(z)$ depends only on one
spatial coordinate. We can impose additional constraints on the
susy parameters $\epsilon_0$, which is compatible with what we have
already imposed.  We then find the corrections to the Lagrangian and SUSY transformation,
which needs on several undetermined parameters.   Depending on the value of
these parameters,  the number of preserved supersymmetry would be 8, 4, or 2.

We impose on the ten-dimensional Majorana Weyl spinor $\epsilon$, the four
conditions including one in (\ref{condition1}),
\be \Gamma^{3456}\epsilon_0=\epsilon_0\, , \quad \Gamma^{3489}\epsilon_0=
-\epsilon_0\, , \quad   \Gamma^{3597}\epsilon_0=-\epsilon_0\, , \quad
\Gamma^{3678}\epsilon_0=-\epsilon_0\, . \label{1dim8}\ee
As the product of the above four conditions is an identity, there
are only three independent conditions, breaking the supersymmetry
to 1/8th or two supersymmetries.

To cancel the supersymmetric variation (\ref{d0l0}) we add to the Lagrangian
the following three terms:
\ba \CL_1 &=&   i \frac{ \psi^{ \prime} } { 4e^2 } \bar{\Psi} \left[
 - \Gamma^{012}  + \frac{1}{ \sin \psi} ( c_0 \Gamma^{456} - c_1 \Gamma^{489}
 - c_2 \Gamma^{597} - c_3 \Gamma^{678}) \right. \nonumber\\
 &&\qquad\quad\left. - \frac{1}{\cos \psi} ( c_0 \Gamma^{789} - c_1 \Gamma^{567}
 - c_2 \Gamma^{648} - c_3 \Gamma^{459}) \right]  \Psi \, ,  \nn \\
\CL_2 &=& - i  \frac{ 2 \psi^{ \prime} }{e^2 \sin \psi } \Tr (
   c_0  \phi_4 [ \phi_5, \phi_6 ]  - c_1  \phi_4 [ \phi_8, \phi_9]
 - c_2 \phi_5 [ \phi_9, \phi_7] - c_3 \phi_6 [ \phi_7, \phi_8 ]  ) \nn  \\
&& + i  \frac{2 \psi^{ \prime} }{e^2 \cos \psi }  \Tr ( 
  c_0  \phi_7 [ \phi_8, \phi_9 ]    - c_1 \phi_5 [ \phi_6, \phi_7]
 - c_2 \phi_6 [ \phi_4, \phi_8] - c_3 \phi_4 [ \phi_5, \phi_9 ]  )\, ,  \nn   \\ 
\CL_3 &=& \sum_{I=4}^9 r_I \Tr \phi_I^2,
\ea
where 
\be
r_I =  D \psi^{ \prime 2} [ c_I ( c_I+ \sin^2 \psi) \cot \psi
      + (1-c_I) ( 1-c_I+ \cos^2 \psi ) \tan \psi  ]
     - D \psi^{ \prime \prime} ( c_I - \sin^2 \psi ),
\ee
and $c_I$'s are real constants satisfying
\ba
&&c_0 + c_1 + c_2 + c_3 = 1\, , \nonumber \\
&&c_4 = c_0 + c_1,\quad c_5 = c_0 + c_2,\quad c_6 = c_0 + c_3\, ,\nonumber \\
&&c_7 = c_2 + c_3,\quad c_8 = c_1 + c_3,\quad c_9 = c_1 + c_2\, .
\ea
The correction of the supersymmetric transformation is
\be
\delta_1 \Psi = - \psi' ( \cot \psi \Gamma \cdot X
                     - \tan \psi \Gamma \cdot Y )  \Gamma^3 \epsilon\, ,
\ee
where
\ba 
\Gamma \cdot X & \equiv &  c_0 \sum_{a=4,5,6} \Gamma^{a} \phi_a
 + c_1 \sum_{a=4,8,9} \Gamma^a \phi_a + c_2 \sum_{a = 5,9,7} \Gamma^a \phi_a 
 + c_3 \sum_{a = 6,7,8} \Gamma^a \phi_a\, ,    \nn \\
\Gamma \cdot Y & \equiv &  c_0 \sum_{p=7,8,9} \Gamma^p \phi_p
 + c_1 \sum_{p=5,6,7} \Gamma^p \phi_p + c_2  \sum_{p=6,4,8} \Gamma^p \phi_p
 + c_3 \sum_{p=4,5,9} \Gamma^p \phi_p\, . \nonumber \\
&&
\ea
Then the total Lagrangian $\CL_0 + \CL_\theta + \CL_1 + \CL_2 + \CL_3 $
is invariant under the corrected supersymmetric transformation. 
For a generic values of constants $c_I$, the number of supersymmetry is two. 
If two of $c_0,c_1,c_2,c_3$ vanish, it is enhanced to four. If only one of
them is nonvanishing, we will have eight supersymmetries as in the previous
sections.

\subsection{the $\tau(y,z)$ case }  

In our previous work \cite{Kim:2008dj}, we classified all supersymmetric 
theories with compatible intersecting Janus interface, both in the two-dimensional and 
three-dimensional case. One can work out a similar analysis to the current case 
with the nontrivial $\theta$ term.
Let us first focus on the case where the coupling constants are
functions of two coordinates $e^2(y,z)$ and $\theta(y,z)$. 
As before, compatible supersymmetric conditions can be expressed by introducing
a constant Majorana Weyl spinor $\epsilon_0$ defined by
\be
 \epsilon(y,z) = e^{ -\frac12 \psi(y,z) \Gamma^{0123} } \epsilon _0\, .
\ee
One can impose three compatible supersymmetry conditions
\begin{align}
\Gamma^{2789} \epsilon_0 & = \Gamma^{3456} \epsilon_0 = \epsilon_0,  \\
- \Gamma^{2459} \epsilon_0 & = \Gamma^{3456} \epsilon_0 = \epsilon_0, \\
- \Gamma^{2567} \epsilon_0 & = \Gamma^{3456} \epsilon_0 = \epsilon_0  , 
\end{align} 
which implies
\be  
\Gamma^{2648} \epsilon_0 = \Gamma^{3489} \epsilon_0 
  = \Gamma^{3597} \epsilon_0 = \Gamma^{3678} \epsilon_0 = - \epsilon_0.
\ee
Each condition breaks the supersymmetry to 1/2 and imposing the 
three at the same time breaks the supersymmetry to the minimal one 1/16.

With these supersymmetric conditions we consider the following interface 
Lagrangian
\begin{align}
{\cal L}_1 = 
&- i \frac{ \partial_3 \psi}{ 4 e^2} 
  \Tr \left(  \bar{ \Psi} \Gamma^{012} \Psi \right)
+ i \frac{ \partial_2 \psi}{ 4 e^2 } 
  \Tr \left(  \bar{ \Psi} \Gamma^{013} \Psi \right)   \nn \\
& +  i \frac{ \partial_3 \psi}{ 4 e^2 }   \Tr \left( \bar{ \Psi}  
  ( \csc \psi  M_3  - \sec \psi N_3 ) \Psi \right)    
+ i \frac{ \partial_2 \psi}{ 4 e^2 }   \Tr \left( \bar{ \Psi} 
  \left(  \csc \psi  M_ 2 - \sec \psi N_2   \right) \Psi \right),
\end{align}
and
\begin{align}
{ \cal L}_2 = 
& - 2 i \frac{ \partial_3 \psi}{ e^2} \csc \psi  
 \Tr \left( c_0 \phi_4 [ \phi_5, \phi_6] - c_1 \phi_4 [ \phi_8 , \phi_9 ] 
- c_2 \phi_5 [ \phi_9 , \phi_7] - c_3 \phi_6 [ \phi_7 , \phi_8] \right) \nn \\
& + 2 i \frac{ \partial_3 \psi}{ e^2} \sec \psi 
 \Tr  \left( c_0 \phi_7 [ \phi_8, \phi_9] - c_1 \phi_5 [ \phi_6, \phi_7] 
- c_2 \phi_6 [ \phi_4, \phi_8] - c_3 \phi_4 [ \phi_5 , \phi_9 ]  \right) \nn \\
& - 2 i \frac{ \partial_2 \psi}{ e^2} \csc \psi 
 \Tr \left( b_0 \phi_7 [ \phi_8, \phi_9 ] - b_1 \phi_5 [ \phi_6, \phi_7]
- b_2 \phi_6 [ \phi_4 , \phi_8] - b_3 \phi_4 [ \phi_5 , \phi_9 ]  \right)\nn \\
& + 2 i \frac{ \partial_2 \psi }{ e^2} \sec \psi 
 \Tr \left( b_0 \phi_4 [ \phi_5 , \phi_6] - b_1 \phi_4 [ \phi_8, \phi_9] 
- b_2 \phi_5 [ \phi_9, \phi_7] - b_3 \phi_6 [ \phi_7, \phi_8 ]   \right), 
\end{align}
where $M_m,N_m,$ ($m=2,3$) are matrices defined by
\begin{align}
M_3 & \equiv c_0 \Gamma^{456} - c_1 \Gamma^{489} 
              - c_2 \Gamma^{597} - c_3 \Gamma^{678} \nn \\
N_3 & \equiv c_0 \Gamma^{789} - c_1 \Gamma^{567} 
              - c_2 \Gamma^{648} - c_3 \Gamma^{459} \nn \\
M_2 & \equiv b_0 \Gamma^{789} - b_1 \Gamma^{567} - b_2 \Gamma^{648}
              - b_3 \Gamma^{459} \nn \\
N_2 & \equiv -  \left(  b_0 \Gamma^{456} - b_1 \Gamma^{489}
              - b_2 \Gamma^{597} - b_3 \Gamma^{678} \right),
\end{align}
and $c_i$, $b_i$ are real parameters satisfying 
\be
\sum_{ i = 0}^3 c_i = \sum_{ i = 0 }^3 b_i = 1,
\ee
so that there are six independent parameters. For convenience
we also denote $c_a^{(2)} = b_a$ and $c_a^{(3)} = c_a$.
Note that the following properties hold for $M_m$ and $N_m$:
\begin{align}
\Gamma^{0123} M_m \epsilon & = N_m \epsilon ,   \nn \\ 
\left( \cos \psi M_m + \sin \psi N_m \right) \epsilon & = \Gamma^m \epsilon . 
\end{align}

Now we define the correction to the supersymmetric transformation \eqref{susy0}
$\delta_1 \Psi$ as
\be
\delta_1 \Psi =  - \partial_3 \psi B_3 \Gamma^3 \epsilon 
         - \partial_2 \psi B_2 \Gamma^2 \epsilon ,
\ee
where 
\begin{equation} 
B_m = \cot \psi \sum_{ a=4}^9 c_a^{(m)} \Gamma^a \phi_a
      - \tan \psi \sum_{ a= 4}^9 c_{a+3}^{(m)} \Gamma^a \phi_a, 
  \quad (m=2,3),
\end{equation}
In this expression $c_4^{(m)},\ldots,c_9^{(m)}$ are given
in terms of $c_1^{(m)},c_2^{(m)},c_3^{(m)}$ as
\begin{align}
&c_4 \equiv c_0 + c_1, \quad c_5 \equiv c_0 + c_2, 
                        \quad c_6 \equiv c_0 + c_3,  \nn \\
&c_7 \equiv c_2 + c_3 ,\quad c_8 \equiv c_1 + c_3 ,\quad c_9 \equiv c_1 + c_2,
\end{align}
and
\begin{align}
& b_4 \equiv b_2 + b_3, \quad b_5 \equiv b_1 + b_3,
                        \quad b_6 \equiv b_1 + b_2 , \nn \\
& b_7 \equiv b_0 + b_1,\quad b_8 \equiv b_0 + b_2, \quad b_9 \equiv b_0 + b_3.
\end{align}
The index $a$ is understood to be cyclic in $4,5,\ldots,9$, i.e., 
$c_{10} = c_4$, $c_{11} = c_5$, and so on. With these, it is not difficult
to show that
\begin{equation}
( \cos \psi M_m  - \sin \psi N_m  ) B_n  \epsilon  = 
 \left[ - B_n + 2 \sum_{ a = 4}^9 c^{(m)}_a ( c^{(n)}_a \cot \psi
 - c^{(n)}_{ a+3} \tan \psi ) \Gamma^a \phi_a \right]  \Gamma^m \epsilon.
\end{equation}
Also note that for$ a= 4,5,\ldots, 9$,
\begin{align}
c_a + c_{ a+3} &= 1, \nonumber \\
c_a b_{ a+3} - b_a c_{ a+3}  &= c_a - b_a, \nonumber \\
c_{ a+3} - b_{ a+3} &= - ( c_a - b_a) .
\end{align}

We are ready to find the supersymmetric Langrangian. First, one can
show that the zeroth variation 
$\delta_0(\mathcal{L}_0 + \mathcal{L}_1 + \mathcal{L}_2)$ is cancelled by
the first order terms from $\delta_1\mathcal{L}_1$. The only nonvanishing
terms are $\delta_1\mathcal{L}_1$ and the second order terms from 
$\delta_1\mathcal{L}_0$, which should be cancelled by introducing another
term $\mathcal{L}_3$ in the Lagrangian. Utilizing the above properties among
the parameters and matrices, it can readily be shown that the desired
term is
\begin{align}
\CL_3 =& - \sum_{ m= 2, 3}  \frac{ ( \partial_m \psi)^2}{e^2} \sum_{ a=4}^9  
\left[ c^{(m)}_a (1+ c^{(m)}_a \csc^2 \psi ) 
 + c^{(m)} _{ a+3} ( 1+ c^{(m)}_{ a+3} \sec^2 \psi ) \right] \phi_a^2 \nn \\
& +  \sum_{ m= 2,3} \frac{ \partial_m^2 \psi}{ 2 e^2}  
 \sum_{ a=4}^9 \left[ c_a^{(m)} \cot \psi - c_{a+3}^{(m)} \tan \psi \right] 
               \phi_a^2 \nn  \\
& - \frac{ \partial_2 \psi \partial_3 \psi}{ e^2} ( \csc^2 \psi - \sec^2 \psi )
\sum_{ a= 4, 5, 6} ( c_a - b_a ) \Tr ( \phi_a \phi_{ a+3} ) \nn \\
& +  \frac{ \partial_2 \partial_3 \psi}{ e^2} ( \cot \psi + \tan \psi ) 
\sum_{ a=4,5,6}  ( c_a - b_a) \Tr  ( \phi_a \phi_{ a+3}  ) .
\end{align}
Then the full Lagrangian $\CL = \CL_0 + \CL_\theta + \CL_1 + \CL_2 + \CL_3$ 
is invariant under the suupersymmetric transformation
$(\delta_0 + \delta_1) \CL = 0.$

\subsection{the $\tau(x,y,z)$ case }  
When the coupling constants depend on all three coordinates, $e^2(x,y,z)$
and $\theta(x,y,z)$, there are two independent supersymmetry conditions:
\begin{align}
& \Gamma^{1467} \epsilon_0 = \Gamma^{2475} \epsilon_0 
          = \Gamma^{3456} \epsilon_0 = \epsilon_0, \nn \\
&  \Gamma^{1458} \epsilon_0 = \Gamma^{2468} \epsilon_0 
          = \Gamma^{3478} \epsilon_0 = \epsilon_0 .
\end{align}
As in the previous section, we define
\begin{align}
M_1 \equiv a_1 \Gamma^{467} + a_2 \Gamma^{458} , 
         &\quad N_1 \equiv a_1 \Gamma^{589} + a_2 \Gamma^{679}, \nn \\
M_2 \equiv b_1 \Gamma^{475} + b_2 \Gamma^{468} , 
         &\quad N_2 \equiv b_1 \Gamma^{689} + b_2 \Gamma^{597}, \nn \\
M_3 \equiv c_1 \Gamma^{456} + c_2 \Gamma^{478} , 
         &\quad N_3 \equiv c_1 \Gamma^{789} + c_2 \Gamma^{569},
\end{align}
where parameters satisfy the relation
\begin{equation}
 a_1 + a_2 = b_1 + b_2 = c_1 + c_2 = 1.  
\end{equation}
The correction to the supersymmetric transformation is
\begin{equation}
\delta_1 \lambda = - \partial_1 \psi B_1 \Gamma^1 \epsilon
                   - \partial_2 \psi B_2 \Gamma^2 \epsilon 
                   - \partial_3 \psi B_3 \Gamma^3 \epsilon, 
\end{equation}
where $B_m$ ($m=1,2,3$) are given by
\begin{equation}
B_m = \cot \psi \left( \Gamma^4 \phi_4
 +  \sum_{ i=5}^8 c^{(m)}_i \Gamma^i \phi_i  \right)
 - \tan \psi \left( \Gamma^9 \phi_9
 + \sum_{ i = 5}^8 ( 1 - c^{(m)}_{ i } )  \Gamma^i \phi_i \right),
\end{equation}
and here we introduced the notations as before,
\begin{align}
& a_5 = a_2, \quad a_6 = a_1, \quad a_7 = a_1, \quad a_8 = a_2, \nn \\
& b_5 = b_1, \quad b_6 = b_2, \quad b_7 = b_1, \quad b_8 = b_2, \nn \\
& c_5 = c_1, \quad c_6 = c_1, \quad c_7 = c_2, \quad c_8= c_2, \nn \\
& c^{(1)}_i \equiv a_i, \quad c^{(2)}_i \equiv b_i, \quad c^{(3)}_i \equiv c_i
   \qquad ( i =  5,6,7,8).
\end{align}
Then the correction terms to the Lagrangian again turn out to consist of
three terms:
\begin{align}
\CL_1 =&
 - \frac{i}{4e^2} \Tr \big[ 
   \partial_1\psi \bar\lambda \Gamma^{023} \lambda 
 + \partial_2\psi \bar\lambda \Gamma^{031} \lambda 
 + \partial_3 \psi \bar\lambda \Gamma^{012} \lambda \nn \\
&\qquad\qquad
 - \partial_1 \psi \bar\lambda ( \csc \psi M_1 - \sec \psi N_1 )\lambda \nn \\ 
&\qquad\qquad
 - \partial_2 \psi \bar\lambda ( \csc \psi M_2 - \sec \psi N_2 )\lambda \nn \\
&\qquad\qquad
 - \partial_3 \psi \bar\lambda ( \csc \psi M_3 - \sec \psi N_3 )\lambda \big], 
\end{align}
\begin{align}
\CL_2 =& - 2 i \frac{\partial_1 \psi}{e^2} \Tr \left\{ 
    \csc \psi ( a_1 \phi_4 [ \phi_6, \phi_7] + a_2 \phi_4 [\phi_5, \phi_8])
  - \sec \psi ( a_1 \phi_5 [ \phi_8, \phi_9] + a_2 \phi_6 [\phi_7, \phi_9])
   \right\} \nn \\
& - 2 i \frac{\partial_2 \psi}{e^2} \Tr \left\{
    \csc \psi ( b_1 \phi_4 [ \phi_7, \phi_5 ] + b_2 \phi_4 [ \phi_6, \phi_8 ])
  - \sec \psi ( b_1 \phi_6 [ \phi_8, \phi_9] + b_2 \phi_5 [ \phi_9, \phi_7])
   \right\} \nn \\
& - 2 i \frac{\partial_3 \psi}{ e^2} \Tr \left\{
    \csc \psi ( c_1 \phi_4 [ \phi_5, \phi_6] + c_2 \phi_4 [ \phi_7, \phi_8] )
  - \sec \psi ( c_1 \phi_7 [ \phi_8, \phi_9] + c_2 \phi_5 [ \phi_6, \phi_9] )
   \right\},
\end{align}
and
\begin{align}
\CL_3 =& - \sum_{m = 1}^3  \frac{ ( \partial_m \psi)^2 }{ 2 e^2 }
  \Tr \left[ ( 1 + \csc^2 \psi ) \phi_4^2 + ( 1 + \sec^2 \psi) \phi_9^2 \right] 
     \nn \\
& - \sum_{m = 1}^3 \frac{ ( \partial_m \psi )^2 }{ 2 e^2 } 
    \sum_{ i = 5}^8 \Tr \left[ 1 + ( c_i^{(m)})^2 \csc^2 \psi
                + ( 1 - c_i^{(m)} )^2 \sec^2 \psi \right] \phi_i^2 \nn \\ 
& + \sum_{m = 1}^3 \frac{ \partial_m^2 \psi }{ 2 e^2 } 
  \Tr\left[ \cot\psi \left( \phi_4^2 + \sum_{i = 5}^8 c_i^{(m)} \phi_i^2\right]
 - \tan\psi \left( \phi_9^2 + \sum_{i = 5}^8 ( 1-c_i^{(m)} )^2 \phi_i^2 \right)
           \right]  \nn \\
& - \frac1{e^2}\left[
            \partial_1 \psi \partial_2 \psi ( \csc^2 \psi - \sec^2 \psi)
          - \partial_1 \partial_2 \psi ( \cot \psi + \tan \psi ) \right] \nn\\
&\qquad\qquad\qquad\qquad \times 
  \Tr\left[ (a_2 - b_1) \phi_5 \phi_6 + (a_1 -b_1) \phi_7 \phi_8 \right] \nn\\
& - \frac1{e^2}\left[
            \partial_2 \psi \partial_3 \psi ( \csc^2 \psi - \sec^2 \psi)
          - \partial_2 \partial_3 \psi ( \cot \psi + \tan \psi ) \right] \nn\\
&\qquad\qquad\qquad\qquad \times 
  \Tr\left[ (b_2 - c_1) \phi_6 \phi_7 + (b_1 -c_1) \phi_5 \phi_8 \right] \nn\\
& - \frac1{e^2}\left[
            \partial_3 \psi \partial_1 \psi ( \csc^2 \psi - \sec^2 \psi)
          - \partial_3 \partial_1 \psi ( \cot \psi + \tan \psi ) \right] \nn\\
&\qquad\qquad\qquad\qquad \times 
  \Tr\left[ (c_2 - a_1) \phi_5 \phi_7 + (c_1 -a_1) \phi_6 \phi_8 \right].
\end{align}
It is straightforward to show that the full Lagrangian is invariant,
\begin{equation}
 ( \delta_0 + \delta_1)( \CL_0 + \CL_\theta + \CL_1 + \CL_2 + \CL_3 ) = 0 .
\end{equation}

\noindent{\bf \large Acknowledgment}

CK and KML are supported in part by the KOSEF SRC Program through CQUeST Sogang
University. KML is supported in part by KRF-2007-C008 program. CK is supported
in part by WCU Program of MEST with Grant No.\ R32-2008-000-10130-0. 
EK is supported in part by KRF-2005-084-C00003, EU FP6 Marie Curie Research \& 
Training Networks MRTN-CT-2004-512194 and HPRN-CT-2006-035863 through 
MOST/KICOS.

\end{document}